\documentclass[
aps,pra,
reprint,
a4paper,
amsmath,
longbibliography,
]{revtex4-1}

\usepackage[graphicx]{realboxes}
\usepackage{booktabs}
\usepackage{siunitx}
\usepackage[]{hyperref}
\usepackage{multirow}
\usepackage{bbm}
\usepackage{bbold} 
\usepackage{float}

\newcommand{\dis}{\mathcal{D}}

\newcommand{\be}{\begin{eqnarray}}
\newcommand{\ee}{\end{eqnarray}}
\newcommand{\ket}[1]{\ensuremath{\left| {#1} \right>}}
\newcommand{\bra}[1]{\ensuremath{\left< {#1} \right|}}
\newcommand{\ketS}[1]{\ensuremath{| {#1} \rangle}}
\newcommand{\braS}[1]{\ensuremath{\langle {#1}|}}

\newcommand{\ketbra}[2]{\ensuremath{| {#1} \rangle \langle {#2} |}}

\newcommand{\caf}{\ensuremath{^{40}{\rm Ca}^{+}\, }}

\newcommand{\create}{\ensuremath{{\,\hat{a}^{\dagger}}}}
\newcommand{\destroy}{\ensuremath{\,\hat{a}}}

\usepackage{etoolbox}
	\newtoggle{arXiv}
	\toggletrue{arXiv}

\newcommand{\Dis}[1]{\hat{\mathcal{D}}(#1)}

\newcommand{\expect}[1]{$\langle #1 \rangle$}
\newcommand{\lket}[1]{\ket{#1}_L}

\newcommand{\sigx}{\hat{X}}
\newcommand{\sigy}{\hat{Y}}
\newcommand{\sigz}{\hat{Z}}
\newcommand{\seq}{seq 0}
\newcommand{\Sxl}{\hat{S}_{x}}
\newcommand{\Szl}{\hat{S}_{z}}

\newcommand{\Sil}{\hat{S}_{i}}

\newcommand{\sigxl}{\hat{X}_{L}}
\newcommand{\sigyl}{\hat{Y}_{L}}
\newcommand{\sigzl}{\hat{Z}_{L}}
\newcommand{\sigil}{\hat{\sigma}_{L}^j}

\newcommand{\ino}{\psi_{\text{in}}}

\usepackage{tikz}
\usepackage{quantumcircuits}

\begin{document}

\title{Encoding a qubit in a trapped-ion mechanical oscillator}

\author{C. Fl{\"u}hmann}\email{Corresponding author, christaf@phys.ethz.ch}
\author{T. L. Nguyen}
\author{M. Marinelli}
\author{ V. Negnevitsky}
\author{K. Mehta}
\author{J. P. Home} \email{Corresponding author, jhome@phys.ethz.ch}

\affiliation{Institute for Quantum Electronics, ETH Z\"urich, Otto-Stern-Weg 1, 8093 Z\"urich, Switzerland}

\date{\today}
\begin{abstract}
The stable operation of quantum computers will rely on error-correction, in which single quantum bits of information are stored redundantly in the Hilbert space of a larger system. Such encoded qubits are commonly based on arrays of many physical qubits, but can also be realized using a single higher-dimensional quantum system, such as a harmonic oscillator \cite{97Chuang,16Marios}. A powerful encoding is formed from a periodically spaced superposition of position eigenstates \cite{01Gottesman, 17Albert, 18Noh}. Various proposals have been made for realizing approximations to such states, but these have thus far remained out of reach \cite{02Travaglione, 06Pirandola2,10Vasconcelos, 16Weigand, 17Keith}. Here, we demonstrate such an encoded qubit using a superposition of displaced squeezed states of the harmonic motion of a single trapped \caf ion, controlling and measuring the oscillator through coupling to an ancilliary internal-state qubit \cite{18Fluhmann}.
We prepare and reconstruct logical states with an average square fidelity of $87.3 \pm 0.7 \%$, and demonstrate a universal logical single qubit gate set which we analyze using process tomography. For Pauli gates we reach process fidelities of $\approx 97\%$, while for continuous rotations we use gate teleportation achieving fidelities of $\approx 89 \%$. The control demonstrated opens a route for exploring continuous variable error-correction as well as hybrid quantum information schemes using both discrete and continuous variables \cite{15Andersen}. The code states also have direct applications in quantum sensing, allowing simultaneous measurement of small displacements in both position and momentum \cite{17Terhal,1955VanNeumann}.
\end{abstract}


\maketitle

The basic unit of quantum information is the qubit. Such a two-state system can be stored in corresponding physical systems, like the spin of an electron. However in practice, the need to correct inevitable errors requires qubits to be stored in physical systems of higher dimension where the larger state space allows detection of errors without disturbing the stored logical information \cite{Devitt13}. Typically, such a larger space is provided by the collective space of multiple physical qubits. Operations acting on the full state space are required in order to perform encoding, measurement and logical control \cite{Devitt13}. An alternative approach is to use a single higher dimensional quantum system, such as a harmonic oscillator or a cavity field mode \cite{97Chuang,16Marios}. The use of a single system requires less resources and offers simplified control, which for microwave cavities has allowed demonstrations of logical qubits encoded and manipulated using so called ``cat'' codes \cite{Heeres2017}. These codes are designed for correcting photon loss, which enabled extension of qubit coherence using error-correction by feedback \cite{16Ofek}. An alternative oscillator code has been proposed by Gottesmann, Kitaev and Preskill (GKP) \cite{01Gottesman}, which is based on sets of displacements generating a periodic grid with a unit cell area of $2 h$ in real phase-space. Compared to other oscillator codes this encoding has been shown to offer the highest correction performance, even outperforming the cat code for the photon loss channel \cite{17Albert}.  Code states can be realized by multi-component superpositions of displaced squeezed states. These states are challenging to engineer and their preparation requires non-linear couplings \cite{01Gottesman}. Once the qubit is encoded,
full quantum state control can be achieved combining relatively simple Gaussian transformations with measurements of the oscillator \cite{01Gottesman}. 

In this Letter, we experimentally demonstrate encoding, logical readout and full control of a GKP qubit in a trapped \caf ion motional oscillator. We generate the code states, and measure the spatial as well as the momentum probability densities, revealing their 2D periodic grid-like non-local structure in phase space. The preparation of these ``grid'' oscillator states is based on coupling the oscillator to two atomic pseudo-spin states $\ket{0} \equiv \ket{S_{1/2}, m_j=1/2}$ and $\ket{1} \equiv \ket{D_{5/2}, m_j=3/2}$ via state-dependent optical forces (SDF) \cite{04Haljan}, combined with post-selected internal-state readout \cite{18Fluhmann}. This toolbox also allows us to  read out the encoded qubit state, and by combining sets of such measurements we perform encoded state tomography. We then extend these tools using direct oscillator displacements and atomic-state rotations to implement and characterize a universal single logical qubit gate set, including non-Clifford gates teleported onto the code.

\begin{figure*}[tb]
	\resizebox{1\textwidth}{!}{\includegraphics{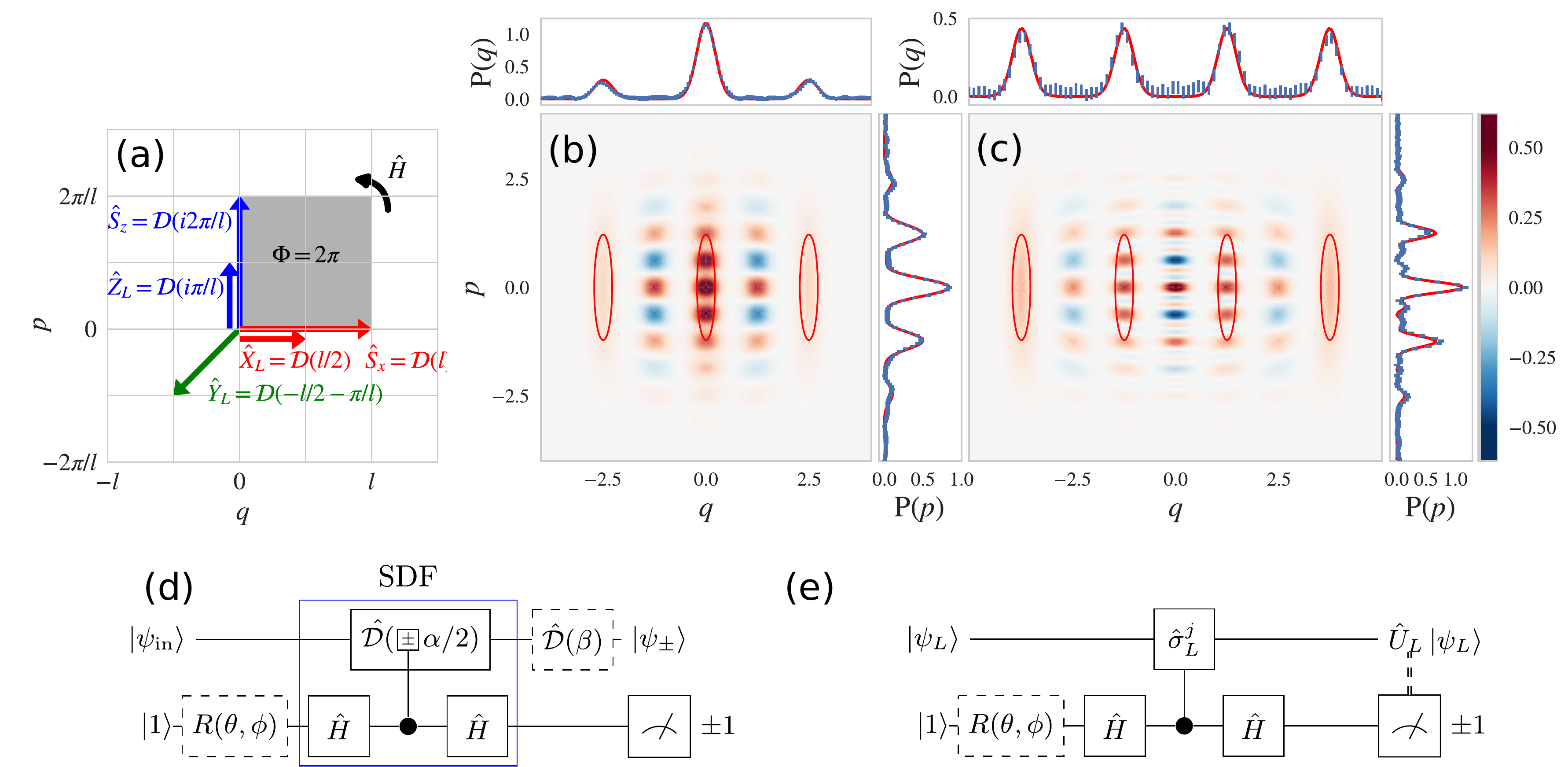}}
	\caption{Grid state encoding and control. (a) Phase-space displacements for the two stabilizer operators $\Sxl$, $\Szl$ and the Pauli operators, showing also the relevant phase space area $\Phi$. We define dimensionless position $\hat{q}$ and momentum $\hat{p}$ such that for a coherent oscillator state: $\braS{\alpha}\hat{q}\ketS{\alpha}=\text{Re}(\alpha)$, $\braS{\alpha}\hat{p}\ketS{\alpha}=\text{Im}(\alpha)$ (see SI). The Hadamard $\hat{H}$ operation is given by a $\pi /2$-rotation of phase space. Approximate grid code states: (b) encoding using the non zero coefficients $c_1=c_{-1}=c_0/2$ and displaying the $\lket{0}$ state and (c) encoding with $c_{-2}=c_{-1}=c_0=c_1$ and showing $\lket{1}$. In each case we plot the Wigner-function of the grid code states obtained by Lindblad-master equation simulation of the experiment (SI). The red ellipses show the position and r.m.s size of the displaced squeezed wave-packets building up these grid states. The two marginal distributions $P(p), P(q)$ are plotted at the corresponding sides of the central figure. Here the curve obtained by simulation is shown in red while blue points with error bars present measurement results (SI) \cite{Wallentowitz95}. (d) Circuit used for grid state qubit control. Operations inside the blue box are an equivalent circuit for the SDF laser pulse. Performed together with the internal-state readout this realizes the modular measurement. The additional dashed-box operations with $\alpha=l_j/2 = 2\beta$ allow teleportation of gates, which is shown in the circuit given in (e) and implementing the continuous operation $\hat{U}_L^j(\theta,\phi)$.
}
	\label{fig:States}
\end{figure*}

In the stabilizer formalism a qubit code subspace is defined within a higher-dimensional Hilbert space by the action of a set of mutually commuting stabilizer operators \cite{Devitt13}. These form the error-check operators which are measured in order to detect logical qubit errors. Such measurements should not disturb the stored information, therefore it is required that the stabilizer operators also commute with the generators of the qubit subspace, given by the Pauli operators. For a single harmonic oscillator, Gottesmann, Kitaev and Preskill showed that a set of stabilizer and Pauli operations can be constructed from displacements in the oscillator phase space. The operator for a displacement is $\Dis{\alpha}=e^{\alpha\hat a^{\dagger} -\alpha^*\hat a}$, where $\alpha$ is a complex number giving the size and direction of the displacement and $\create, \destroy$ are the creation and annihilation operators of the oscillator  \cite{05Schleich}. Displacement operators are in general non-commutative, following $[\Dis{\alpha}, \Dis{\beta}] = 2 i e^{i\Phi} \sin(\Phi) \Dis{\alpha} \Dis{\beta}$ with $\Phi=\text{Im}(\beta\alpha^*)$. Displacements satisfying $\Phi = k \pi, k\in\mathbbm{Z}$ commute, while for $\Phi = (2 k + 1)\pi/2$ they anti-commute. It follows that $\sigxl \equiv \Dis{l/2}$, $\sigzl \equiv  \Dis{i\pi/l}$ and $\sigyl \equiv  \Dis{-l/2-i\pi/l}$, will commute with the stabilizer operators $\Sxl \equiv  \Dis{l}$, $\Szl \equiv  \Dis{i2\pi/l}$, because $\Phi=\pi$ or $0$, while the two stabilizer operators have $\Phi=2\pi$ and thus also commute.  $\Phi$ is independent of the parameter $l$, which can thus be varied. Figure \ref{fig:States} (a) gives a summary of the logical operators in phase-space. The simultaneous eigenstates of $\sigzl, \Szl, \Sxl$ are the computational basis states and are periodic with respect to the three phase-space shifts. The displacement operators are non-Hermitian $\Dis{\alpha}^\dagger=\Dis{-\alpha}$, nevertheless  the action of $\sigxl,\sigyl,\sigzl$ and their Hermitian conjugates are identical on the periodic code states. This ensures the correct behavior for Pauli operators (for more details, see supplemental information (SI)).

The computational basis states consist of an infinite array of position eigenstates \cite{01Gottesman}. These ideal code states are unphysical, since they cannot be normalized. Approximations to these states are given by finite superpositions of displaced squeezed states \cite{01Gottesman}:
\be
\lket{0}= \sum_{k \in \mathbb{Z}}^{\pm|k_{\rm max}|} c_k \Dis{k l}\ket{r}, \lket{1} = \Dis{l/2}\lket{0}
\label{gkpform}
\ee
where $\ket{r}=\hat{S}(r)\ket{0}$ is a squeezed vacuum state with the squeezed axis aligned with position. $\hat{S}(r) = e^{r (\destroy^2 - \create^2)/2}$ is the corresponding phase space squeezing operator, where we define the squeezing parameter $r$ to be real and positive. The weight of the displaced components is given by the real pre-factors $c_k$.
This approximate form approaches the ideal states for larger $r$ and $k_{\rm max}$ \cite{06Glancy, 01Gottesman}.

\begin{figure*}[tb]
	\resizebox{1\textwidth}{!}{\includegraphics{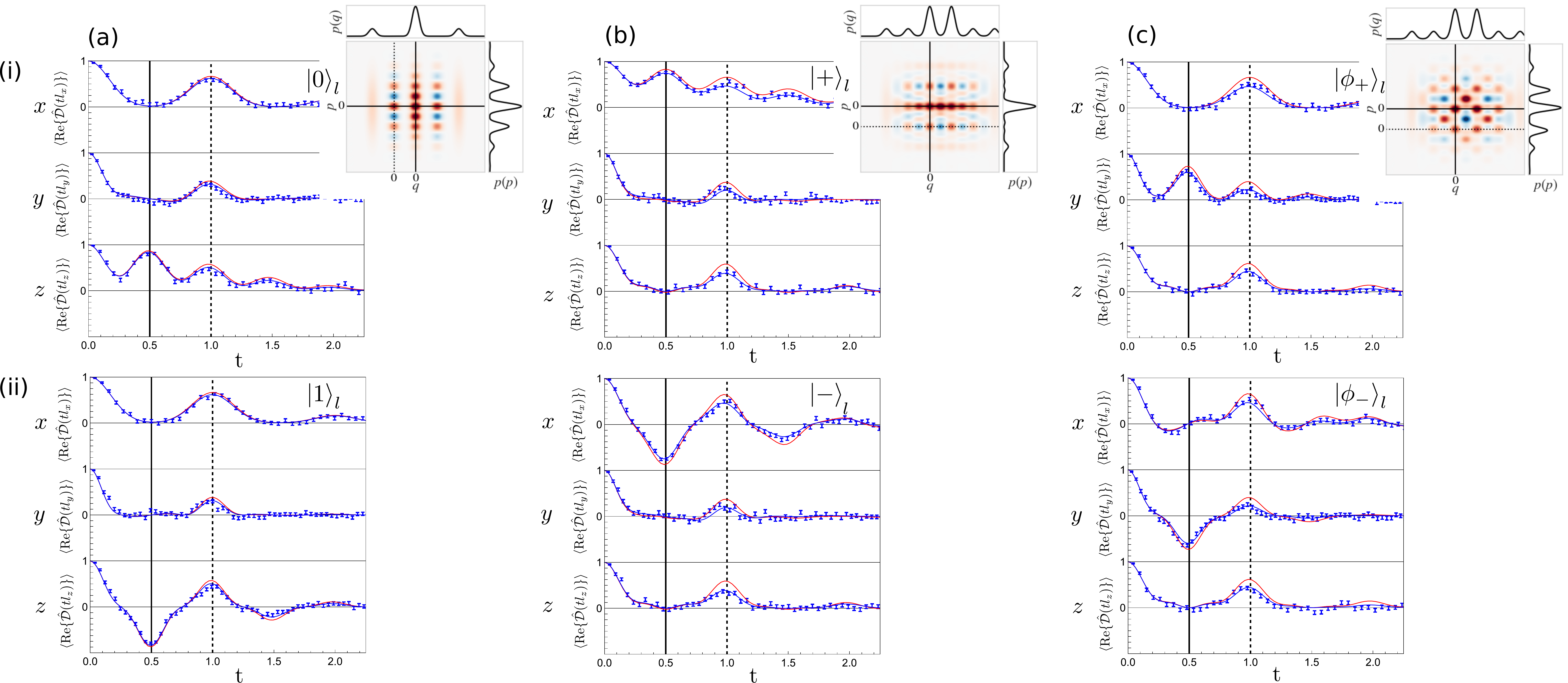}}
	\caption{Logical readout. Column (a) $\sigzl$ eigenstates, (b) $\sigxl$ eigenstates and (c) $\sigyl$ eigenstates. For all data settings were close to $r \approx 0.9$, $l \approx \sqrt{2\pi}$ and we set $c_1=c_{-1}=c_0/2$. The $\sigzl$ eigenstates are three-component superpositions while in (b) and (c) these are six-component superpositions given by coherent superpositions of the $\sigzl$ eigenstates components. Each state is analyzed by sweeping the displacement paramter $\alpha = t l_j$ along three phase space axis: $l_x = l$ (top), $l_y = -l_x - l_z$ (middle), $l_z = 2 \pi i/l $ (bottom) and $t$ is real and parametrizes these directions. The readout for $t=0.5$ (solid vertical line) corresponds to a $\langle\sigil\rangle$ Pauli readout, while $t=1$ (dashed line) corresponds to stabilizer read out $\langle\Sil\rangle$. Blue points represent measured data with the error bars given as standard errors of the mean (SEM). The red line shows an analytic calculation for the chosen $l$, $r$ and $c_k$  while the blue line is a Lindblad-master simulation including motional dephasing. The insets show theoretical Wigner functions of relevant motional states. In each case the -1 eigenstate (row (ii)) is prepared from +1 eigenstate (row (i)) via the application of a Pauli operation, thus the Wigner-function of the -1 state has the same qualitative pattern but is shifted in phase space. This is accounted for in the plot by the shift of the relevant plot axes. Solid (dotted) lines are valid for +1 (-1) eigenstates.
	}
	\label{fig:Readout}
\end{figure*}

Experimentally, we create the code states in the axial motional mode of a single \caf ion at close to  $\omega_{\rm{m}} \approx 2 \pi \times 1.85 \rm{\:MHz}$ by starting from a squeezed vacuum state produced by reservoir engineering \cite{15Kienzler}, followed by repeated application of a modular variable measurement sequence \cite{16Kienzler, 02Travaglione, 18Fluhmann, 15Lo}. A modular measurement is implemented by a two stage process shown in the circuit in figure \ref{fig:States} (d) excluding operations inside the dashed boxes.  First i) the ancillary internal-state qubit in $\ket{1}$ is coupled to the oscillator using a resonant internal state-dependent force implementing $\Dis{\alpha/2\sigx}$ with $\sigx$ the internal-state Pauli operator (Blue box in figure \ref{fig:States} (d)). This is followed by ii) detection of the internal-state of the ion by resonance fluorescence. Conditioned on detecting no scattered photons, this implements the operator $\hat{E}_{+}=1/2(\Dis{\alpha/2}+\Dis{-\alpha/2})$ on the oscillator, producing a superposition of two displaced copies of the initial state. This process is repeated with appropriately chosen $\alpha$ to build up multi-component superpositions of the form given in equation \ref{gkpform}.
Figure \ref{fig:States} (b) and (c) show Wigner function simulations (SI) as well as measured probability densities (SI) \cite{Wallentowitz95} in both position $\text{P}(q)=\text{Tr}(\hat{q}\hat{\rho})$ and momentum $\text{P}(p)=\text{Tr}(\hat{p}\hat{\rho})$ for two approximate code eigenstates created using two rounds of modular variable measurement. Part (b) shows an example of a $\lket{0}$ state consisting of three displaced components created using measurement displacements $\alpha_1 =\alpha_2 = l$, while part (c) presents a four component version of $\lket{1}$ created using $\alpha_1 = l$, $\alpha_2=2l$. In both cases we choose $l\approx\sqrt{2\pi}$ which generates code states with similar modularities in $q$ and $p$. To minimize sensitivity to motional dephasing a low phonon number is desirable. Therefore states with a symmetric extent in $q$ and $p$ and with high weight close to the origin are preferable. This favors $r\approx 0.9\approx 7.8 \rm{\:dB}$ and the 3-component state on which we base our encoding in the results below. The marginals of the simulated Wigner functions provide a theory curve for the measured probability densities and match well with the data presented in figure \ref{fig:States} (b) and (c).

In order to create eigenstates of the other Pauli operators as well as implement arbitrary logical control, we perform two types of operation on the logical states. Pauli operations are simple displacements $\Dis{\alpha}$. These are experimentally implemented by applying an oscillating voltage resonant with the trap frequency $\omega_{\rm{m}}$ to one of our trapping electrodes \cite{96Leibfried}. The magnitude $|\alpha|$ of the displacement is set by the product of the pulse amplitude and duration, and the direction $\arg(\alpha)$ by the phase of the oscillating
tone. For continuous operations, we use a modification of the modular variable measurement, which involves the full circuit shown in figure \ref{fig:States} (d).  We use ancilla controlled displacements  $\dis(l_j/4 \hat{X} )$ along variable directions $j=x,y,z$ defined using $l_x = l$, $l_z = 2 \pi i/l$ or $l_y = -l_x - l_z$. This together with an unconditional corrective displacement $\dis(l_j/4)$, which ensures to remain within the code space, realizes the controlled logical Pauli operation shown in figure \ref{fig:States} (e). Prior to the application of these pulses, a rotation  $\hat{R}(\theta,\phi)=\cos(\theta/2)\mathbb{1}+i \sin (\theta/2)(\sin(\phi)\sigx +\cos(\phi)\sigy) $ is performed on the ancilliary internal-state qubit using a resonant Rabi oscillation. Conditional on the dark detection event this full circuit implements $\hat{U}_L^j(\theta,\phi)=\sqrt{2}(\cos(\theta/2)\ketbra{+_j}{+_j}_{L}+\sin(\theta/2)e^{i\phi}\ketbra{-_j}{-_j}_{L})$ on the oscillator state, where $\ket{\pm_j}_L$ denotes the $\pm 1$ ideal eigenstate of $\hat{\sigma}_L^j$. In the context of error-correction codes, this transformation is often referred to as teleportation of the gate onto the code \cite{97Knill2}. This operation is only unitary if we set $\theta=\pi/2$, resulting in a rotation $\hat{U}_L^j = e^{i\phi/2} R_{L}^j (\phi) = e^{i\phi/2} \text{exp}(-i\phi  \sigil /2)$ around the $j$-axis of the Bloch sphere \cite{BkNielsen}.
However in state preparation it is not necessary to perform unitary operations as long as the pure target state is reached, and thus we used values of $\theta = 0, \pi, \pi/2$ in the state preparations presented in figure \ref{fig:Readout}.

Outcomes of the modular variable measurements are used to read out the logical operators and stabilizers. The internal-state readout performed in the modular measurement sequence has outcome probabilities  $P(1) = \bra{\psi_{\rm in}} \hat{E}_+^{\dagger}\hat{E}_+ \ket{\psi_{\rm in}}$ and $P(-1)=1-P(1)$ with the expectation value of the readout given by \expect{\sigz} $= P(1) - P(-1) = \bra{\ino}\hat{Q}\ket{\ino}$ where $\hat{Q}=\text{Re}\{\Dis{\alpha}\}$ and $\ket{\psi_{\rm in}}$ is the initial oscillator state. For appropriately chosen complex displacements this circuit realizes readout of the logical operators, which require only the real value since they are Hermitian for ideal code states. The readout probabilities are dependent on interference of the two displaced copies of the original state, which depends both on the state overlaps and on geometric phases \cite{18Fluhmann} (SI).

\begin{figure*}[tb]
	\resizebox{1\textwidth}{!}{\includegraphics{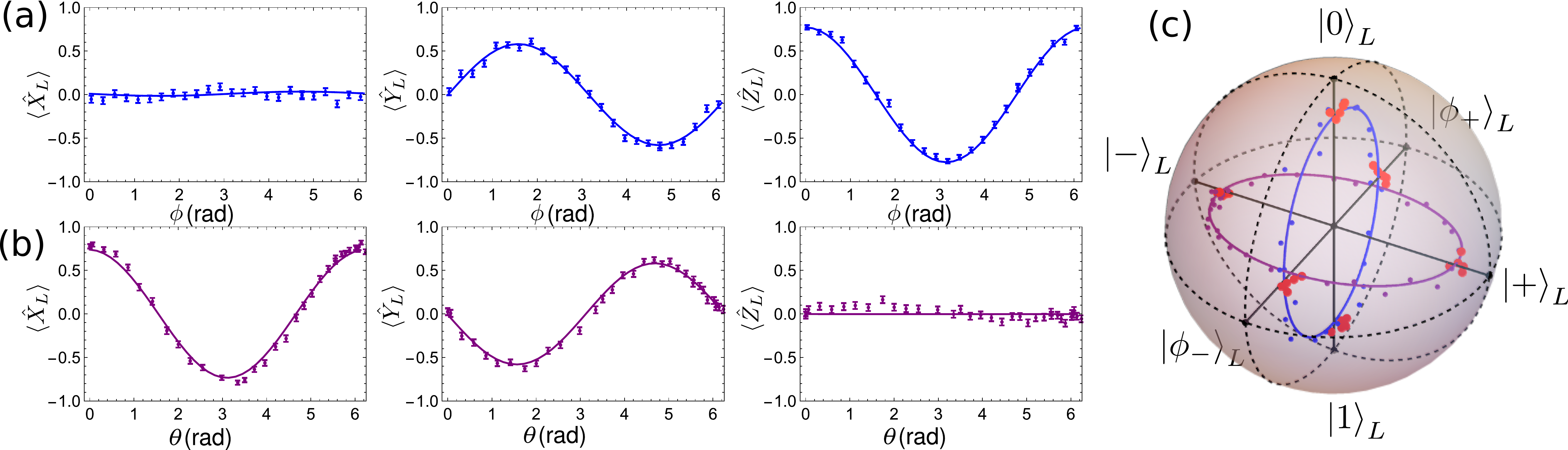}}
	\caption{Arbitrary single qubit operations (a) $\hat{U}_L^x(\pi/2,\phi)\lket{0}$ realizing a rotation around the $x$-axis of the Bloch sphere. (b) $\hat{U}_L^x(\theta,\pi/2)\lket{0}$. In (a) and (b) $\ket{\psi_\text{in}}=\lket{0}\propto ( \Dis{-l} + 2 + \Dis{l} )\ket{0.91}$ with $l$=2.36 and errors are given as SEM while the solid line show the Lindblad master simulation of the experiment. (c) Summary of the continuously varied operations presented in (a) and (b) on the Bloch sphere. In addition as red points the states shown in figure $\ref{fig:Readout}$ and all the process tomography input states are shown.
}
	\label{fig:Operations}
\end{figure*}

Figure \ref{fig:Readout} shows the results of such measurements performed on each of the eigenstates of $\sigxl$, $\sigyl$, $\sigzl$ with non-zero coefficients $c_1=c_{-1}=c_0/2$. Each is prepared by first preparing the three-component $\lket{0}$, followed by appropriate displacements and teleported operations (for all settings see SI). Figure \ref{fig:Readout} shows additionally theoretical Wigner function plots of the created motional states. For each initialized state, we give the readout as a function of the displacement amplitude along three directions, which are parametrized as $\alpha = t l_j$ with the real number $t$.  The periodic nature of the code states can be observed in the experimental data (blue points) presented in figure \ref{fig:Readout}. Also shown are theory curves (red lines) together with a master equation simulations which includes the effects of motional dephasing (blue line). We again see good agreement of the latter with our measurement.
For $t = 1$ we measure the stabilizer operators for which we find an average over the six input states $\langle\Sxl\rangle=56 \pm 1 \% \: (65.8\%)$, $\langle\Szl\rangle=41 \pm 1 \% \:(59.2\%)$ the value given in the bracket is the expectation due to the approximate nature of the code states (red line).
At $t = 0.5$ the measurement reads-out the logical Pauli operators which allows us to reconstruct the logical qubit density matrix $\hat{\rho}_L=\frac{1}{2}(\mathbbm{1} +\langle \sigxl \rangle \sigxl + \langle \sigyl \rangle \sigyl + \langle \sigzl \rangle \sigzl)$. We quantify the logical qubit quality by calculating the fidelity between the reconstructed state and the ideal state $\ket{\text{id}}$ as $F(\hat{\rho}_L,\ket{\text{id}})= \bra{\text{id}} \hat{\rho}_L \ket{\text{id}}$. The average state creation and readout fidelity of five data sets measured over several days was $87.3\%$ with a standard deviation between the averages of $0.7\%$. In this case the finite approximation limits the achievable average fidelity to $90.8\%$. All the measured states are shown as red points on the Bloch sphere in figure \ref{fig:Operations} (c).

Readout of logical operators allows us to examine the teleported gates $\hat{U}_L^j(\theta,\phi)$ implemented using the ancilla qubit. First we set $\theta = \pi/2$ and use the controlled $\sigxl$ operation. This implements a rotation around the $x$-axis $\hat{U}_L^x(\pi/2,\phi)=R_{L}^x (\phi)$, with the choice of phase $\phi$ denoting the rotation angle. This operation applied to $\lket{0}$ is shown in figure \ref{fig:Operations} (a). We see that the value of $\langle \sigxl \rangle$ is largely unaffected, while a clear rotation is seen in the $\langle \sigyl \rangle$ and $\langle \sigzl \rangle$ signals. Figure \ref{fig:Operations} (b) shows similar data, obtained using $\phi = \pi/2$ while varying the value of $\theta$ again using the input state $\lket{0}$ and the controlled $\sigxl$ operation. Although this transformation is useful in state preparation it is not unitary. The states produced using both operations are shown on the Bloch sphere in figure \ref{fig:Operations} (c).

\begin{figure*}[tb]
	\resizebox{1\textwidth}{!}{\includegraphics{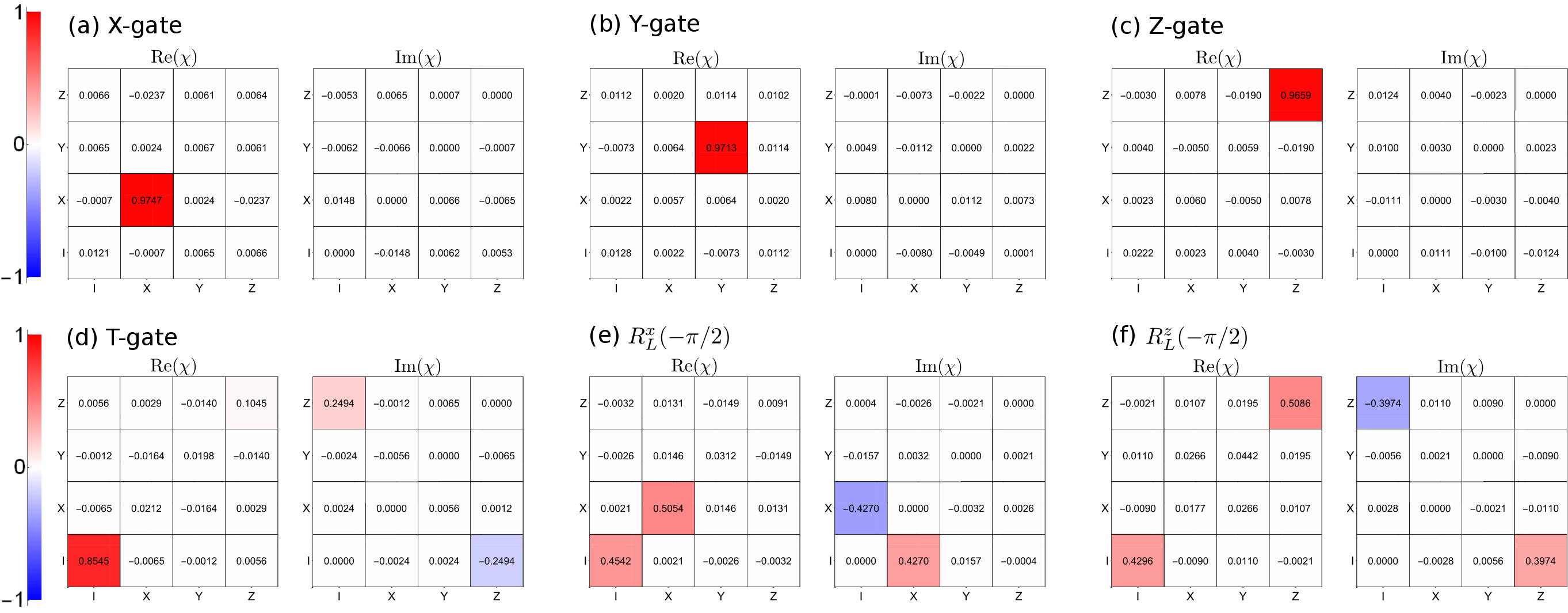}}
	\caption{Process tomography of logical operations, characterized by the $\chi$ matrix. The reconstructed operations are (a)-(c) Pauli gates, implemented by phase space displacements, (d) $\hat{T}_L = R_L^z(\pi/4)$, (e) $R_L^z(-\pi/2)$ and (f) $R_L^z(-\pi/2)$ rotations implemented by teleportation using the internal-state qubit as an ancilla. The Hadamard operation $\hat{H}_L$ is given by an oscillator phase space rotation of $\pi/2$, which can be conveniently realized as an update to the readout direction in the classical control system. Such a permutation of readout results trivially leads to a perfect process tomography result.
}
	\label{fig:Proc}
\end{figure*}

To characterize the performance of our qubit operations we use quantum process tomography. The six approximate eigenstates of $\sigxl$, $\sigyl$, $\sigzl$ are used as input states, which are then subjected to the process of interest. The input density matrix $\hat{\rho}_L^{\text{in}}$ as well as the output density matrix $\hat{\rho}_L^{\text{out}}$ are reconstructed by readout in the three Pauli bases. The process can be described by a linear map $\mathcal{E}(\hat{\rho}_L^{\text{in}})=\hat{\rho}_L^{\text{out}}=\sum_{mn} \hat{\sigma}_L^m \hat{\rho}_L^{\text{in}} \hat{\sigma}_L^n \hat{\chi}_{mn}$, which is fully defined by the complex matrix $\chi$. From the measurement results, we obtain $\chi$ using a constrained least square optimization of the set of linear equations relating input to output states (SI). Results for a universal set of logical gates are shown in figure \ref{fig:Proc}. The presented set of gates is given by all three Pauli operations $\sigxl$, $\sigyl$, $\sigzl$, together with the $\hat{T}_L$ gate ($R_L^z(\pi/4)$), and two $\pi/2$-rotations $R_L^x(-\pi/2)$ and $R_L^z(-\pi/2)$ about orthogonal axes. The latter three were performed by gate teleportation.

The quality of each of these operations can be evaluated by calculating the process fidelity $F_{\hat{O}}=\text{Tr}(\chi\chi_\text{id})$ between the experimentally obtained  $\chi$ and the ideal logical qubit matrix $\chi_\text{id}$. For the Pauli operations we find $\text{F}_{\sigil}=97 \%$ while for the three reconstructed partial rotations around the $z$ and $x$-axis we find $\text{F}_{\hat{T}_L}=\text{F}_{R_L^z(\pi/4)}=92 \%$, $\text{F}_{R_L^x(-\pi/2)}=91 \%$, $\text{F}_{R_L^z(-\pi/2)}=87 \%$. The numerical optimization used for determining $\chi$ makes the evaluation of error bars non-trivial and we thus forego quoting error bars here. Additionally we note that the Hadamard gate can be implemented as an update of the readout directions. By definition, the process tomography routine then gives an ideal process matrix and a process fidelity of 1.

The evaluation of the logical qubit gates given above is agnostic to its physical realization. Although the process tomography analysis captures the relevant information about storing and manipulating quantum information, the underlying quantum states are those of the oscillator. In practice, the change in the states of the oscillator produces a number of complications which require further study. In the work above, this includes that teleported gates generate additional grid state components, and the change in the state adjusts the logical readout levels. Sequences of Pauli gates given by phase space shifts can be optimized in order to maintain population at low $\bar{n}$ which minimizes dephasing.

Extensions to this work would include performing error-correction or control of multiple encoded qubits. Multi-qubit gates can be performed for example using lasers to couple two oscillation modes of one ion mediated via the internal-states. Local modes of different ions can be coupled using the Coulomb interaction \cite{11Brown, 11Harlander}. Laser and Coulomb coupling allow for the realization of the beam-splitter operation, which can also be used for error-correction \cite{06Glancy}. An alternative realization of error correction is through phase estimation using modular variable measurements \cite{16Weigand}. Phase estimation is currently challenging due to the destructive nature of a bright detection, which would require post-selection. This might be alleviated in the future by using integrated cavities, which allow to reduce photon recoil through directional absorption and emission. 
To reach fault tolerance, an improvement in the approximation to ideal grid states would be required. Theoretical results indicate that levels of $r\sim 14.8$~dB might be required when the current methods are concatenated \cite{18Fukui2}. Protocols harnessing the analog as well as the discrete character of the grid qubits may relax these requirements to $r  \sim 10 \rm{\:dB}$, which is within the range of our current experimental system \cite{18Fukui2, 15Kienzler}.
The demonstrated control provides a new route for investigating quantum error correction, while opening up the possibility to realize improved sensing of small phase-space displacements in both position and momentum \cite{17Terhal}.

\begin{acknowledgments}
We thank Daniel Kienzler, Ludwig de Clercq and Hsiang-Yu Lo for important contributions to the apparatus. We acknowledge support from the Swiss National Science Foundation through the National Centre of Competence in Research for Quantum Science and Technology (QSIT) grant $51NF40-160591$. We acknowledge support from the Swiss National Science Foundation under grant no. $200020\_165555/1$. KM is supported by an ETH Z\"urich Postdoctoral Fellowship $17-1-FEL-50$. The research is partly based upon work supported by the Office of the Director of National Intelligence (ODNI), Intelligence Advanced Research Projects Activity (IARPA), via the U.S. Army Research Office grant $W911NF-16-1-0070$. The views and conclusions contained herein are those of the authors and should not be interpreted as necessarily representing the official policies or endorsements, either expressed or implied, of the ODNI, IARPA, or the U.S. Government. The U.S. Government is authorized to reproduce and distribute reprints for Governmental purposes notwithstanding any copyright annotation thereon. Any opinions, findings, and conclusions or recommendations expressed in this material are those of the author(s) and do not necessarily reflect the view of the U.S. Army Research Office.
\end{acknowledgments}

\textbf{Author Contributions:} Experimental data were taken and analyzed by CF, using an apparatus with significant contributions from VN, MM, CF, TL, KM. The paper was written by CF and JPH, with input from all authors. Experiments were conceived by CF and JPH.
The authors declare that they have no competing financial interests.

\iftoggle{arXiv}{
}{\bibliography{./myrefs2}}

\iftoggle{arXiv}{

\iftoggle{arXiv}{
	\clearpage
	{\LARGE\bfseries Supplementary Information\\}
}{}

\section{dimensionless position and momentum}
We choose definitions of dimensionless position and momentum such that we have a simple connection to phase space: $\hat{q}=\sqrt{\frac{m\omega}{2\hbar}}\hat{q}_r$ and $\hat{p}=\sqrt{\frac{1}{2 m\omega \hbar}}\hat{p}_r$ here $\hat{q}_r, \hat{p}_r$ are the real-space position and momentum operators. Using these definitions we find  $\braS{\alpha}\hat{q}\ketS{\alpha}=\text{Re}(\alpha)$, $\braS{\alpha}\hat{p}\ketS{\alpha}=\text{Im}(\alpha)$ and $[\hat{q},\hat{p}]=i/2$. This definition simplifies working with position, momentum and displacement operators simultaneously.

\section{Logical operations}
We have argued in the main text that the action of the logical Pauli operator becomes Hermitian when acting on the ideal code states which still holds approximately for approximate code states \cite{16Ketterer}. Further the stabilizer operators act as the identity operation on the code states. Therefore we have:
\begin{align}
&\Dis{l/2}=\sigxl\approx \sigxl^{\dagger}=\Dis{l/2}^\dagger=\Dis{-l/2}\\\notag
&\Dis{i\pi /l}=\sigzl\approx \sigzl^{\dagger}=\Dis{i\pi /l}^\dagger=\Dis{-i\pi/l}\\\notag
&\Dis{-l/2-i\pi /l}=\sigyl\approx \sigyl^{\dagger}=\Dis{l/2+i\pi /l}\\\notag
&\Sxl =\Dis{l} \approx \mathbb{1}_l\\\notag
&\Szl =\Dis{2i\pi/ l}\approx \mathbb{1}_l
\end{align}

The logical Pauli operations should fulfill the relation $\hat{\sigma}_i \hat{\sigma}_j = \delta_{ij} \mathbb{1} + i\epsilon_{ijk}\sigma_k$ here $\delta_{ij}$ is the Kronecker delta and $\epsilon_{ijk}$ the Levi-Civita symbol. From this relation then the usual Pauli commutation and anti-commutation relations follow.
We find:
\begin{align}
&\sigxl^2=\Dis{l/2}^2=\Dis{l}=\Sxl \approx \mathbb{1}_l \\\notag
&\sigyl^2=\Dis{-l/2-i\pi /l}^2=\Dis{-l-i\pi}=\Sxl^\dagger\Szl^\dagger \approx\mathbb{1}_l\\\notag
&\sigzl^2= \Dis{i\pi /l}^2= \Dis{2i\pi /l}=\Szl \approx\mathbb{1}_l\\\notag
&\sigxl\sigyl=\Dis{l/2}\Dis{-l/2-i\pi /l}=i\Dis{-i\pi /l}=i\sigzl^\dagger \approx i\sigzl \\\notag
&\sigxl\sigzl=\Dis{l/2}\Dis{i\pi /l}=-i\Dis{l/2+i\pi /l}=-i\sigyl^\dagger  \approx-i\sigyl \\\notag
&\sigyl\sigzl=\Dis{-l/2-i\pi /l}\Dis{i\pi /l}= i\Dis{-l/2}= i\sigxl^\dagger \approx i\sigxl
\end{align}

\section{Phase-space control}
The presented experiments rely on excellent control of the oscillator phase space. This requires a stable and well calibrated motional frequency. Additionally we need to be able to reference the orientation of the squeezed state, to the state-dependent force displacement direction and to the unconditional displacements direction implemented by the oscillating drive to a trapping electrode.

\subsection{Motional frequency calibration}
We calibrate the motional frequency of $1.85 \rm{\:MHz}$ with a Lorenzian line-with of approximately $10 \rm{\:Hz}$ to an accuracy of around $10 \rm{\:Hz}$. Experimentally frequency drifts of up to $1.8 \rm{\:Hz/min}$ are observed thus we recalibrate the motional frequency every $5 \rm{\:min}$. A quick, accurate and robust frequency calibration is thus required. This is achieved by first ground state cooling of the ions motion and then applying the oscillating voltage with frequency $\omega_{\rm{m}}+ \delta$ to one of our trapping electrodes. This is followed by a wait time of around 4 ms after which we apply a second oscillating voltage with opposite phase to the first pulse. The final motional state is then probed with a red sideband pulse. In case of $\delta=0$ the motional state should return to the ground state after this sequence and thus the red sideband will not be able to change the internal-states. If there is a detuning $\delta$ present then the ion ends in an excited motional state, and the spin can be inverted by the red sideband pulse. Using a squeezed initial oscillator state and a squeezed basis probe pulse gives a geometrical advantage for this calibration. Nevertheless the method using the ground state proved to be experimentally more robust and was therefore used.

\subsection{Calibration of displacement directions}
The SDF direction and the squeezed state orientation are both defined by the difference phase of the red and blue sideband laser. The creation of the squeezed state and the SDF pulse use the exact same electronic and optical signal paths thus their directions stay fixed with respect to each other. In order to match the unconditional displacement direction to this direction we create a squeezed oscillator state which we first displace using the SDF laser pulse. (Addition of a carrier $\pi/2$-rotation before and after the SDF pulse allows the implementation of $\Dis{\alpha\hat{Z}}$ \cite{18Fluhmann}). Then we aim to invert this displacement using the tickling (oscillating drive to the trapping electrode) pulse. Finally we probe the created oscillator state using the squeezed basis analog of the red sideband \cite{16Kienzler}. In case we succeed to invert the unconditional displacement due to the laser we are not able to flop the qubit. We exploit the squeezed state features in this calibration by displacing along the squeezed axis in order to calibrate the duration of the oscillating voltage and along the anti-squeezed axis in order to find its phase.

\subsection{calibration sequence used for grid state experiments}
Before each grid state experiment we calibrate the squeezing parameter $r$ as well as the SDF coupling strength by two independent measurement \cite{15Kienzler,18Fluhmann}. Then we match the tickling voltage displacement to the SDF as described above. Finally we calibrate the angles and phases of the used ancilla carrier rotations.

\section{Measured probability densities}
We have a report in preparation explaining in detail how we experimentally access the marginal distributions $\text{P}(q)=\text{Tr}(\hat{q}\hat{\rho})$ and $\text{P}(p)=\text{Tr}(\hat{p}\hat{\rho})$ of the Wigner functions presented in figure \ref{fig:States} (b) and (c). The basic idea is theoretically described in \cite{Wallentowitz95,12Casanova}. Using a $\pi/2$-rotation before the modular measurement allows readout of $\langle\text{Im}\{\Dis{\alpha}\}\rangle$, this together with the modular measurement reading out $\langle\text{Re}\{\Dis{\alpha}\}\rangle$ gives full information about the symmetric characteristic function $\chi(\alpha)=\langle\Dis{\alpha}\rangle$. The characteristic function is the Fourier transform of the Wigner-function. Thus by measuring $\chi(t)$ $(\chi(it) )$ varying the real parameter $t$ and performing a discrete Fourier transform we will obtain $\text{P}(p)$ ($\text{P}(q)$). We zero-pad our data for the discrete Fourier transform and perform bootstrapping in order to obtain error bars. See the public data repository for the directly measured values of the characteristic function.

\section{Logical state readout}
The readout is based on the non-commutativity of phase space displacements $\Dis{\alpha}\Dis{\beta}=e^{2 i \text{Im}(\alpha \beta^*)} \Dis{\beta}\Dis{\alpha}$ and interference of the grid states with themselves. This we explain in the following for the $l_z$ readout direction and on a $\ket{\ino}=\lket{0}= \sum_k c_k \Dis{k l}\ket{r}$ state. The modular measurement probability is given by $P(\pm1)=\bra{\ino} \hat{E}_\pm^{\dagger}\hat{E}_\pm \ket{\ino}$, which is given by the overlap of the un-normalized post-measurement state of the oscillator $\hat{E}_\pm \ket{\ino}$ with itself. This post-measurement state consists of two displaced copies of the input state $\lket{0}$.  One copy is displaced down the momentum axis by $\alpha/2$ while the other copy is displaced up. Only components originating from the same initial grid state component $k$ i.e. $\Dis{kl}\ket{r}$ have significant overlap. For these overlaps the non-commutativity of the displacement operators lead to different phase factors. Using these relations we find $P(+1)\approx 1/2\left(1 + \sum_k \cos(2 l_z t k l) \bra{r}\Dis{i l_z t k l}\ket{r} \right)$ from which $P(-1)=1-P(+1)$ follows. We can observe this functional form shown at the bottom in figure \ref{fig:fig:Readout} column (a) row (i). The initial increase of $t$ leads first to different phases between the various terms in the sum and the readout signal drops. At $t=0.5$, which corresponds to readout of $\sigzl$ the terms have all phases of multiples as $2\pi$ and ideally would completely rephase. Due to the finite squeezing the overlap characteristic function $O=\bra{r}\Dis{t l_z}\ket{r} =\text{exp}(-|t l_z|^2 e^{-2r})$ reduces which leads to a smaller revival. Note that for a squeezed vacuum state $\ket{r}$, the characteristic function is always real, which is in general not the case. A next such revival occurs at $t=1$ corresponding to readout of the stabilizer operator, where we see that the overlap dropped even further. Using the $\lket{1}$ as an input state due to the different position of the squeezed components the geometric phase factors do change by $2\pi t$ which leads to a negative revival at $t=0.5$.
Similar arguments can be made for the readouts in the other directions for example in the $l_x$ case components originating from neighboring Grid state components will overlap and these overlaps will add up constructively or destructively dependent on the initial relative phase between these components. Here revivals will be reduced due to the finite number of displaced components.

\section{State tomography}
We reconstruct logical grid state qubit states via readout in the three logical Pauli basis $\hat{\rho}_L=\frac{1}{2}(\mathbbm{1} + \langle \sigxl \rangle  \sigxl+ \langle \sigyl \rangle \sigyl + \langle \sigzl \rangle \sigzl)$. Any measurement will yield an expectation value in the range [-1,1] and thus to a valid density operator but with this method we are not able to detect if we left the logical code space.\\
As seen in the previous section the logical readout levels are limited by the underlying approximate code states. In particular they depend on how well the states are eigenstates of the two stabilizer operators $\Sxl$, $\Szl$. The readout of $\Sxl$, $\sigxl$ improves with a higher number of squeezed components. While more initial squeezing improves the $\Szl$ and $\sigzl$ readouts. We optimized the code states in such a way that readouts of $\sigzl$ and $\sigxl$ are limited at a similar level. In turn the $\sigyl$ readout is limited by both the number of components and the squeezing and is thus expected to be lower.


\section{Logical state creation}
In table \ref{Rtable} all pulse sequences used to create the eigenstates of $\sigxl$, $\sigyl$, $\sigzl$ are summarized. Each of the state creation sequences starts with the preparation of a squeezed vacuum state $\ket{r}=\hat{S}(r)\ket{0}$ with $r\approx 0.9$ corresponding to $\approx 7.8 \rm{\:dB}$ of squeezing \cite{15Kienzler}. The squeezed state serves then as the input to a sequence of two modular measurements (Mod) \cite{18Fluhmann}. In almost all preparations we make the two sequential modular measurements with identical displacements given by $\text{Re}(\Dis{\approx \sqrt{2\pi}})$ implemented by $\approx 38 \rm{\:us}$ of SDF laser pulse. This creates the desired $\lket{0}$ with $c_1=c_{-1}=c_0/2$ and any other coefficient $c_k=0$. The other states are created from $\lket{0}$ using appropriate Pauli and teleported gates. For the data presented in figure 2 we used Pauli operations in order to transform from the +1 eigenstate to the -1 eigenstate, while in performing process tomography we created the -1 eigenstates directly from $\lket{0}$ using gate teleportation. After each fluorescence readout we proceed to the next measurement conditional on the dark measurement result otherwise we restart the experimental sequence. This choice is made using real time decisions implemented by an FPGA. This leads to $p \approx 3/8 $ success probability to create $\lket{0}$ and $\lket{1}$ and $p \approx 3/16$ for the other 4 states.

\begin {table}[ !htbp ]%
\label {Rtable} %
\begin {tabular}{ lll }
State 			& Figure 					 																		& Creation pulse sequence\\ 																		
$\lket{0}$ 	& 1 (b)																		& (\seq) $\equiv$ Squeezed pumping \cite{15Kienzler},\\
&&Mod(l), Mod(l)	 \\			
												
$\lket{1}$ 	& 1 (c) 																	& Squeezed pumping \cite{15Kienzler}, Mod(l), Mod(2l)	\\					
																	
$\lket{0}$ 	& 2 (a)	(i)																& (\seq) 	\\																											
																																																										
$\lket{1}$ 	& 2 (a)	(ii)																& (\seq), $\sigxl=\Dis{l/2}$	 \\

$\lket{+}$ 	& 2 (b)	(i)																& (\seq), $\hat{U}_L^x(0,\text{arb.})$	\\													

$\lket{-}$ 	& 2 (b)	(ii)																	& (\seq), $\hat{U}_L^x(0,\text{arb.})$, $\sigzl=\Dis{\pi/l}$	\\ 		

$\lket{\phi_+}$ 	& 2 (c)	(i)													& (\seq), $\hat{U}_L^x(\pi/2,\pi/2)$ \\														

$\lket{\phi_-}$ 	& 2 (c)	(ii)														& (\seq), $\hat{U}_L^x(\pi/2,\text{arb.})$, $\sigzl=\Dis{\pi/l}$\\	 	
$\lket{0}$ 	& 4, $\hat{\rho}_L^{\text{in}}$									& (\seq) \\
																																																																																						
$\lket{1}$ 	& 4, $\hat{\rho}_L^{\text{in}}$																		& (\seq), $\sigxl=\Dis{l/2}$	\\																		

$\lket{+}$ 	& 4, $\hat{\rho}_L^{\text{in}}$																	& (\seq), $\hat{U}_L^x(0,\text{arb.})$ \\														
					
$\lket{-}$ 	& 4, $\hat{\rho}_L^{\text{in}}$			 														& (\seq), $\hat{U}_L^x(\pi,\text{arb.})$\\												

$\lket{\phi_+}$ & 4, $\hat{\rho}_L^{\text{in}}$														& (\seq), $\hat{U}_L^x(\pi/2,\pi/2)$	\\															

$\lket{\phi_-}$ 	& 4, $\hat{\rho}_L^{\text{in}}$												& (\seq), $\hat{U}_L^x(\pi/2,-\pi/2)$												
											
\end {tabular}
\caption {Creation of code states: All states used $l\approx 2.5$ while $r\approx 0.9\approx 7.8 \rm{dB}$}
\label {Rtable} %
\end {table}

\section{Extraction of the process matrix}
We create the six input states described in the previous section and reconstruct their state via readout in the three Pauli bases: $\hat{\rho}_L^j = \sum_{k} o_{jk} \hat{\sigma}_k$ here $j$ labels the input state number, $k$ the Pauli basis element and $o_{jk}$ correspond to the readout results (where we added $o_{j0} = 1/2$ for the identity basis element). Then we apply the process of interest to each input state and reconstruct the output state in the same way with $\lambda_{jk}$ the corresponding measurement results. An arbitrary physical process connecting input states to output states can be expressed in the Pauli basis introducing the process matrix $\chi$: $\mathcal{E}(\hat{\rho}_L^{\text{in}})=\hat{\rho}_L^{\text{out}}=\sum_{mn} \hat{\sigma}_L^m \hat{\rho}_L^{\text{in}} \hat{\sigma}_L^n \hat{\chi}_{mn}$. A linear set of equations for the matrix elements of $\chi_{mn}$ connects $o_{jk}$ to $\lambda_{jk}$: $(\stackrel{\rightarrow}{\lambda} = \beta \stackrel{\rightarrow}{\chi})$ with the matrix $\beta$ calculated from the input state measurements $o_{jk}$. The process matrix $\chi$ is Hermitian and non-negative definite. To ensure these properties we parametrize $\chi=\hat{T}^\dagger\hat{T}$ with \\\\
$\hat{T}=
  \begin{pmatrix}
    t_1 & 0 & 0 & 0 \\
    t_5 + i t_6 & t_2 & 0 & 0\\
		t_{11} + i t_{12} & t_7 +i t_8 & t_3 & 0\\
		t_{15} + i t_{16} & t_{13} +i t_{14} & t_9 + i t_{10} &  t_4\\
  \end{pmatrix}
$\\
a tridiagonal matrix \cite{01James}. We find 4 more constraints on the elements of $\hat{T}$ following from trace preservation of the logical process \cite{Bhandari2016}. Ensuring these constraints we find the $t_i$ elements which minimize $|\beta \stackrel{\rightarrow}{\chi}-\stackrel{\rightarrow}{\lambda}|^2$ using the NMinimise function of Mathematica.\\

The presented method of process tomography is independent of the chosen grid state encoding we solely specify the methods of input state creation, state tomography and analyze how well the implemented processes realize logical single qubit operations. Nevertheless we have additional knowledge about our code states and logical readout and see that a number of effects are not accounted for in process tomography. For example the logical $\pi/2$-rotation around the $z$-axis transforms logical $\sigyl$ readouts to $\sigxl$ readouts in this case we expect the readout levels to increase! Even though in this particular case our realization of the logical qubit improves this will lead to infidelities in the process matrix.\\
Rescaling of the readout directions in order to account for such imprecision is not trivial since in the rotation implemented by gate teleportation typically the underlying states change (i.e more squeezed components or spread out squeezed states) which in general also changes the logical readout level.\\



\section{Lindblad-master equation simulation}
We simulate the experimental sequence of pulses together with a Lindblad operator ($\sqrt{\Gamma}(\hat{a}\hat{a}^{\dagger}+\hat{a}^{\dagger}\hat{a})$) with $\Gamma= 7 \: s^{-1}$. This motional dephasing leads to reduced readout levels of the three Pauli operations. All the other discrepancies of the measured data to the expectation is due to the limited precision in the calibrations required to run the experiments.

\section{Data availability}
The data that support the plots within this paper and other findings of this study are available from the following link http://www.tiqi.ethz.ch/publications-and-awards/public-datasets.html . 
\makeatletter
\apptocmd{\thebibliography}{\global\c@NAT@ctr 30\relax}{}{}
\makeatother

\iftoggle{arXiv}{
}{\bibliography{./myrefs2}}
}{}

\end{document}